\documentstyle[preprint,aps]{revtex}
\begin{document}
\preprint{OITS-529}
\draft
\title{CP violation in a multi-Higgs doublet model with
flavor changing neutral
current }
\author{N.G. Deshpande\footnote{e-mail address: desh@oregon.uoregon.edu} and
Xiao-Gang He\footnote{e-mail address: he@bovine.uoregon.edu}}
\address{Institute of Theoretical Science\\
University of Oregon\\
Eugene, OR97403-5203, USA}
\date{Dec. 1993}
\maketitle
\vspace{-1.0cm}
\begin{abstract}
We study CP violation in a multi-Higgs doublet model
based on a $S_3 \times Z_3$
horizontal symmetry where CKM phase is not the principal source of CP
 violation. We consider two mechanisms for CP violation in this model: a)
CP violation due to complex Yukawa couplings; and b) CP violation due to
scalar-pseudoscalar Higgs boson mixings. Both mechanisms can explain
 the observed CP violation in the neutral Kaon system. $\epsilon'/\epsilon$ due
to neutral Higgs
 boson exchange is small in both mechanisms, but charged Higgs boson con-
tributions can be as large as $10^{-3}$ for a), and
$10^{-4}$ for b). CP violation in
 the neutral B
system is, however, quite different from the Minimal Standard
 Model. The neutron Electric Dipole Moment can be as large as the present ex-
perimental bound, and can be used to constrain charged Higgs boson masses.
 The electron
 EDM is one order of magnitude below the experimental bound  in case b) and
smaller in case a).
\end{abstract}
\pacs{}
\newpage
\section
{ Introduction}

The origin of CP violation, fermion masses and mixings are some of the
outstanding problems of particle physics today. The Minimal Standard Model
(MSM) provides no explanations for the observed fermion masses and mixings. It
is
believed that one needs to go beyond the MSM to solve these problems. There is
no satisfactary solution so far. The best one can do at the present is to
reduce the number of free parameters in the theory regarding fermion masses and
mixings. The understanding of CP violation is also very poor. So far CP
violation has only been observed in the neutral kaon system.  In the MSM CP
violation is from the phase in the CKM matrix\cite{km} $V_{KM}$. The model is
consistent with
observations. It is, however,  important to study different processes and other
models of CP violation to better understand the origin of CP violation. In this
paper we will study some details of CP violation in a model proposed by
Ma\cite{ma}. The model is based on a horizontal $S_3\times Z_3$ symmetry. In
this model there are some interesting relations between the fermion masses and
mixings. The free parameters related to the fermion masses and mixings are
reduced. This model also has some
interesting consequences for CP violation\cite{desh,lav}. Many of the
considerations in this model will have applicability to more general
multi-Higgs models.

There are four Higgs doublets in the $S_3\times Z_3$ model. Their interactions
at the tree level mediate neutral flavor changing current. Like any multi-Higgs
doublet models with flavor changing neutral current at the tree level, there
are different mechanisms for CP violation.
CP violation can arise in three places in
this type of models: 1) Non-trivial phase in the $V_{KM}$ matrix; 2)Non-trivial
phases in the flavor changing Yukawa couplings; and 3) Mixings of scalar and
pseudoscalar Higgs bosons. In cases 2) and
3), CP violation can occurs at the tree level by exchanging neutral Higgs
bosons\cite{fcp}.
These models have much richer phenomenology for CP violation than the MSM. In
the MSM, CP violation in the neutral Kaon system can be explained by exchanging
W-bosons at the one loop level (the "box diagram")\cite{lim}. There are
specific predictions for $\epsilon'/\epsilon$ resulting from
the direct CP violation in $K_{L,S}\rightarrow 2\pi$\cite{buras};
and predictions for CP violation in the neutral B system\cite{bcp}. The
Electric Dipole Moment of neutron is generated at the three loop level, and is
less than
$10^{-31} ecm$\cite{he}. The electron EDM is even smaller\cite{krip}. In the
$S_3\times Z_3$ model, the CP violation coming from the phase in the CKM matrix
 is inadequate to account for the $\bar \epsilon$ parameter in the K system.
One of the other two has to be invoked. We consider the consequences of either
mechanism for: (i) $\epsilon'/\epsilon$; (ii) CP violation in the neutral B
system; and (iii) the neutron and electron EDMs. The results are dramatically
different from the MSM.

\section
{Yukawa couplings in the $S_3\times Z_3$ model}

In the $S_3\times Z_3$ model, there are four Higgs doublets, $\phi_{1,2,3,4}$.
The quarks and Higgs bosons transform under the $S_3\times Z_3$ symmetry
as\cite{ma}
\begin{eqnarray}
q_{3L}\;, t_R\;, b_R\;, \phi_1: (1, 1)\;,\nonumber\\
(q_{1L}, q_{2L})\;, (\phi_3, \phi_4):  (2, \omega)\;,\nonumber\\
 (c_R, u_R)\;,s_R, d_R): (2,\omega^2)\;,\nonumber\\
\phi_2: (1,\omega^2)\;,
\end{eqnarray}
where $\omega \neq 1$, $\omega^3 = 1$ is the $Z_3$ element.
The Yukawa couplings consistent with the $S_3\times Z_3$ symmetry are given by
\begin{eqnarray}
L_Y &=& -f_1(\bar q_{1L} \tilde \phi_3 u_R + \bar q_{2L} \tilde \phi_4 c_R)
-f_2 \bar q_{3L}\tilde \phi_1 t_R\nonumber\\
&-& f_3 (\bar q_{1L}\phi_2 s_R + \bar q_{2L} \phi_2 d_R)
-f_4(\bar q_{1L}\phi_3 b_R + \bar q_{2L} \phi_4 b_R)\nonumber\\
&-&f_5(\bar q_{3L}\phi_3 d_R + \bar q_{3L} \phi_4 s_R) -f_6 \bar q_{3L}\phi_1
b_R
+ H.C.
\end{eqnarray}
where $\tilde \phi_i = (\phi^{0*}_i, -\phi_i^-)^T$. Without loss of generality
we work in a basis where all Vaccum Expectation Values (VEV) are real. When the
neutral components develop VEVs, $<\phi_i> = v_i$, we obtain the quark mass
matrices
\begin{eqnarray}
M^u &=&  \hat M^u = \left (\matrix{f_1 v_3&0&0\cr 0&f_1 v_4&0\cr
0&0&f_2v_1}\right )\;,\nonumber\\
\\
M^d &=& \left( \matrix{0&f_3v_2&f_4v_3\cr
f_3v_2&0&f_4v_4\cr f_5v_3&f_5v_4&f_6v_1\cr}\right)\nonumber\;.
\end{eqnarray}
The quark phases can be chosen such that
\begin{eqnarray}
M^d = \left ( \matrix{ 0&a&\xi b\cr a&0&b\cr \xi c&c&d\cr}\right )\;,
\end{eqnarray}
with a, b, c, d real and $\xi = |\xi|e^{i\sigma}$ complex. $M^d$ can be
diagonalized by a bi-unitary transformation
\begin{eqnarray}
M^d = V_L \hat M^d V^\dagger_R\;.
\end{eqnarray}
Here $\hat M^d$ is the diagonalized down quark mass matrix. $V_L$ and $V_R$ are
unitary matrices. Because the up quark mass matrix is already diagonlized, $V_L
$ is the CKM matrix $V_{KM}$.

 It is convenient to work in a basis of the Higgs bosons in which the Goldstone
bosons are removed. To this end we define the
following\cite{lav}
\begin{eqnarray}
\left ( \matrix{\phi_1\cr \phi_2\cr \phi_3\cr \phi_4\cr} \right )
= \left ( \matrix{ {v_1\over v} & {v_2\over v_{12}} & {v_1v_4\over v_{12}
v_{124}} & -{v_1v_3\over vv_{124}}\cr
{v_2\over v} & - {v_1\over v_{12}} & {v_2v_4\over v_{12}v_{124}} &
-{v_2v_3\over v v_{124}}\cr
{v_3\over v} &0&0& {v_{124}\over v}\cr
{v_4\over v} &0& -{v_{12}\over v_{124}}& - {v_3v_4\over v v_{124}}\cr} \right )
\left (\matrix {G\cr H_1\cr H_2\cr H_3\cr} \right )\;,
\end{eqnarray}
where $v_{12}^2 = v_1^2+v_2^2$, $v_{124}^2 = v_1^2 +v_2^2+v_4^2$, and $v^2 =
v_1^2 +v_2^2+v_3^2+v_4^2$.
The transformation is the same for both the neutral and charged Higgs bosons.
For the
neutral Higgs bosons, $G = h^0 + iG_Z$, where $G_Z$ is the Goldstone boson
'eaten'
by Z, and $h^0$ is a physical field whose couplings are the same as the Higgs
boson in the MSM. For the charged Higgs bosons G is the
Goldstone boson 'eaten' by W. In this basis, we have

\begin{eqnarray}
L_Y &=& -(\bar D_L \hat M^d D_R + \bar U_L \hat
M^u U_R)(1 + {Reh^0\over v\sqrt{2}})
\nonumber\\
&-& \bar D_L \tilde Y^d_i D_R {h^0_i\over \sqrt{2}}
- \bar U_L \tilde Y^u_i  U_R
{h^{0*}_i\over \sqrt{2}}\nonumber\\
&-& \bar U_L V_{KM} \tilde Y^d_i D_R h^+_i +
 \bar D_L V^\dagger_{KM} \tilde Y^u_i U_R h^-_i + H.C.\;,
\end{eqnarray}
where $h_i$ are the component fields of $H_i$ with $H_i = (h^+_i,
h^0_i/\sqrt{2})$. $U_{L,R} = (u, c, t)_{L,R}^T$, and $D_{L,R} = (d, s,
b)_{L,R}^T$. The Yukawa couplings are given by
\begin{eqnarray}
\tilde Y^u_1 &=& Diag (0,0, {m_tv_2\over v_{12}v_1})\;,\nonumber\\
\nonumber\\
\tilde Y^u_2 &=& Diag (0, -{m_cv_{12}\over v_4v_{124}}, {m_tv_4\over
v_{12}v_{124}})\;,
\nonumber\\
\nonumber\\
\tilde Y^u_3 &=& Diag ({m_uv_{124}\over v_3v}, -{m_cv_3\over v_{124}v},
-{m_tv_3\over v_{124}v})\;,\nonumber\\
\nonumber\\
\tilde Y^d_1 &=& {1\over v_{12}} V_{KM}^\dagger \left (
\matrix{0&-a{v_1\over v_2}&0\cr -a{v_1\over v_2}&0&0\cr
0&0& d {v_2\over v_1}\cr} \right )V_R\;, \\
\nonumber\\
\tilde Y^d_2 &=&  {1\over
v_{124}}
V_{KM}^\dagger\left ( \matrix {0& a{v_4\over v_3}&0\cr
a{v_4\over v_3}&0&-b{v_{12}\over v_4}\cr
0&-c{v_{12}\over v_4}&d{v_4\over v_{12}}\cr}\right )V_R\;,\nonumber\\
\nonumber\\
\tilde Y^d_3 &=& {1\over v} V_{KM}^\dagger \left (\matrix{0&-a{v_3\over
v_{124}}&
\xi b{v_{124}\over v_3}\cr
-a{v_3\over v_{124}}&0&-b{v_3\over v_{124}}\cr
\xi c {v_{124}\over v_3}& - c{v_3\over v_{124}}& - d{v_3\over
v_{124}}\cr}\right )V_R
\nonumber\;.
\end{eqnarray}

In general $h_i^{0,+}$ are not the mass eigenstates. They will mix with each
other. In particular if CP is violated in the Higgs potential, $Reh_i^0$ and
$Imh^0_i$ will mix. Also
the charged Higgs boson mixing matrix will be complex. We can parametrize the
mixings as
\begin{eqnarray}
\left (\matrix{h^0\cr Reh^0_k\cr Imh^0_k\cr}\right )
&=& \left ( \matrix{\alpha_{00}&\alpha_{0i}&\beta'_{0j}\cr
\alpha_{k0}&\alpha_{ki}&\beta'_{kj}\cr
\alpha'_{k0}&\alpha'_{ki} & \beta_{kj}\cr} \right ) \left (\matrix{R_0\cr
R_i\cr I_j\cr}
\right )\;,\nonumber\\
\\
h^+_i &=& (\gamma_{ij})\eta^+_j\nonumber\;.
\end{eqnarray}
where $R_i$, $I_i$ and $\eta_i$ are the mass eigenstates, the matrix $(\alpha
\beta)$ is a $7\times7$ othogonal matrix, and $(\gamma)$ is a $3\times 3$
unitary matrix.

The specific numbers for $\alpha_{ij}$, $\alpha'_{ij}$, $\beta_{ij}$ and
$\beta'_{ij}$ depend on the details
of the Higgs potential. Unfortunately they are not determined. To simplify the
problem,
we will discuss two cases: a) CP violation only come from complex Yukawa
couplings; and b) CP violation only come from the mixings of real and imaginary
$h^0_i$. Case a) can be realised by constraining certain soft symmetry breaking
terms in the potential\cite{lav}. We further assume, for simplicity, that
$Reh^0_i$ are the mass
eigenstate $R_i$ and consider their effects. The same analysis can be easily
carried out for $Imh^0_i$ in the same way. The source for CP violation is the
non-zero value
for $\sigma$ which is a free parameter.  We will present our
results for $\sigma = 80^0$, which is close to the maximum of the allowed
phase. Case b) can be realised by requiring spontaneous CP violation. The value
of $\sigma$ will be zero and CP violation arises due to scalar-pesudoscalar
Higgs boson mixing. For illustration, we consider the effects of a neutral
mixed state
\begin{eqnarray}
R = cos\theta Reh^0_2 + sin\theta Imh^0_3\;,
\end{eqnarray}
and for the charged Higgs boson we consider mixing
\begin{eqnarray}
\eta^+ = \gamma_{22}h^+_2 + \gamma_{23} h^+_3\;
\end{eqnarray}
where $\gamma_{ij}$ are complex numbers, and
$|\gamma_{22}|^2 +|\gamma_{23}|^2 = 1$.
We assume that $R$ and $\eta^+$ are mass eigenstates,
and all other Higgs bosons
are much heavier and their effects can be neglected.

The details of the results
depend on the specific vules of $\sigma$ in case a) and how Higgs bosons are
mixed in case b). However, the special cases considered here
will serve as a guide for a more complete analysis if the details of the
mixings are known. The general features will be the same. We will comment on
other cases later.

The parameters $a$, $b$, $c$, and $d$ are constrained from the down quark
masses and the CKM mixings. We take as input parameters $a =
0.04 GeV$, $b = 0.25 GeV$, $c= 2.66 GeV$ , $d = 4 GeV$.
The mass eigenvalues for the down quarks are quite insensitive to the phase
$\sigma$. For both cases, we have $m_b =4.8GeV$, $m_s = 149 MeV$ and $m_d = 9.5
MeV$. These values are well within the allowed regions\cite{qm}. The CKM matrix
for case a) is
\begin{eqnarray}
V_{KM} = \left ( \matrix{0.975&-0.222&0.00476\cr
0.221 +i0.0033&0.974+i0.014& 0.043-i0.0015\cr
-0.014+ i 1.2\times 10^{-5}&-0.041-i6.8\times 10^{-4}& 0.998 - i0.034\cr}
\right )\;,
\end{eqnarray}
and for case b)
\begin{eqnarray}
V_{KM} = \left ( \matrix{0.975&0.22&0.0048\cr
-0.219&0.975&-0.0436\cr-0.014&0.0415&0.999}\right)\;.
\end{eqnarray}

The values for the VEV's are not fixed, we only know
$v_3/v_4 = m_u/m_c$. We will use the values in Ref\cite{lav}: $v_1 =v_2 = 44
GeV$, $v_3 = 0.9$ and $v_4 = 238 GeV$ for illustration. We shall comment on
effects of changing these values later.

\section
{ Constraints on the Higgs boson masses from the  neutral K and B meson
systems}

The $S_3\times Z_3$ model has very restrictive allowed values for the
non-trivial CP violating phase in the CKM matrix. The CP violating measure
J\cite{jask} is less than $2.5 \times 10^{-6}$. If CP violation is purely from
the CKM matrix,  it is not possible to explain the observed CP violation in the
neutral K system which requires $J\geq10^{-5}$. Therefore in this model CP
violation due to Higgs boson exchange has to be
considered.
Because the neutral Higgs bosons mediate flavor changing neutral current at the
tree level, there are stringent constraints on their masses
and their interactions arising from experimental data. Some of the best
constraints are from the mass differences in the neutral
K and B systems. We must make sure that the observed CP violation in
the neutral K system $\bar\epsilon = 2.3 \times 10^{-3}e^{i\pi/4}$ is
explained.

 The CP violating parameter $\bar\epsilon$
is given by
\begin{eqnarray}
\bar\epsilon = {ImM_{12}^K\over \sqrt{2}\Delta m_K}e^{i\pi/4}\;,
\end{eqnarray}
where $M_{12}^K$ is the matrix element which mixes $K^0$ with $\bar K^0$, and
$\Delta m_K$ is the mass difference between $m_{K_L}$ and $m_{K_S}$. The
$\Delta S = 2$ Hamiltonian, responsible for $M^K_{12}$, generated by exchanging
neutral Higgs bosons $R_i$ is given by
\begin{eqnarray}
H_{eff} = -{1\over 2M^2_{R_i}}
\left ( \bar d [(\alpha_{ki} +i \alpha'_{ki})\tilde Y^d_{k,12})
{1+\gamma_5\over 2}
+ (\alpha_{ki} -i\alpha'_{ki})\tilde Y^{d*}_{k,21}){1-\gamma_5\over 2}]s\right
)^2\;.
\end{eqnarray}
We obtain
\begin{eqnarray}
M_{12}^K&=& <K^0|H_{eff}|\bar K^0>\nonumber\\
&=& -{f_k^2m_K\over 2 M^2_{R_i}}
( -{5\over 24}{m_K^2\over (m_s+m_d)^2}[(\alpha_{ki} +i \alpha'_{ki})\tilde
Y^d_{k,12})^2
+ (\alpha_{ki} -i\alpha'_{ki})\tilde Y^{d*}_{k,21})^2]
\nonumber\\
&+&(\alpha_{ki} +i\alpha'_{ki})\tilde Y^d_{k,12}(\alpha_{k'i}
-i\alpha'_{k'i})\tilde Y^{d*}_{k',21}
({1\over 12} + {1\over 2} {m_K^2\over (m_s+m_d)^2}) )
\;.
\end{eqnarray}
Here we have used the vaccum saturation and factorization approximation results
for the matrix elements\cite{tram}
\begin{eqnarray}
<K^0|\bar d (1\pm\gamma_5)s \bar d (1\mp\gamma_5)s|\bar K^0> &=&
f_K^2m_K({1\over 6} + {m_K^2\over (m_s+m_d)^2})\;,\nonumber\\
<K^0|\bar d (1\pm \gamma_5)s \bar d (1\pm \gamma_5)s|\bar K^0>&=&
-{5\over 6}f_K^2m_K {m_K^2\over (m_s +m_d)^2}\;.
\end{eqnarray}
The contribution to the mass difference $\Delta m_K$ is given
by $2ReM_{12}$. Similar formula holds for the neutral B system.

To constrain the Higgs boson masses, we require that the neutral Higgs boson
contributions to the mass differences in
the neutral K and B systems to be less than the experimental values:
$\Delta m_K/m_K = 7 \times 10^{-15}$, and $\Delta m_B/m_B = 8 \times 10^{-14}$.
We find that for case a) the tightest constraints on the masses of
$Reh^0_{1,2}$ are from the mass difference $\Delta M_B$ of the neutral  B
mesons which gives $M_{h_1} >2.9 TeV$ and $ M_{h_2}>3.1 TeV$. With these
masses, $Reh^0_{1,2}$ can
not produce large enough
$\bar \epsilon$. Similar consideration yields $M_{h3} > 3.5 TeV$, and we find
the experimental value of $\bar\epsilon$ can now be
produced if the mass is about  $5.6 TeV$. The mass difference
$\Delta M_K$ of the neutral K mesons gives weaker
bounds in all cases.
For case b),  the
experimental value of $\Delta M_B$ constrains $M_R > 3 TeV$. From the
experimental value of $\bar\epsilon$, we obtain $sin\theta
cos\theta/M^2_R = 1.1\times 10^{-8} $ which implies $M_R < 7 TeV$.

Similar analyses for case a) and case b) have been carried out previously in
Ref.\cite{lav} and Ref.\cite{desh} respectively. Our analysis for case a) is
similar to that in Ref.\cite{lav}. We used different values for the paramters
$a, b, c$, and $d$ which obtain smaller mass for $m_b$ ($4.8 GeV$) compared
with the value $5.4 GeV$  used in Ref.\cite{lav}. The analysis for case b) is
different than that in Ref.\cite{desh}. In Ref.\cite{desh}, the averaged
effects of Higgs boson exchange were considered.

{}From the above we see that the Higgs boson masses are constrained to be in
the multi
 TeV region. One would wonder if such heavy Higgs bosons may violate the
unitarity
bound. However this is not the case.
It should be noted that the unitarity bound only apply to $h^0$
Higgs boson mass. Its mass is not constrained in the cases we are considering,
$h^0$ can be light. In the above discussions we have
neglected mixings between $h^0_i$ with $h^0$. If such mixings are large, the
unitarity bound can constrain the $h^0_i$ masses.
However there is enough freedom in our model to make the mixings with $h_0$
sufficiently small
such that the unitarity bounds are always satisfied\cite{desh}. This argument
applies to many models. A well known example is the minimal supersymmetric
standard model. In this model, when the
soft supersymmetry breaking parameter goes to infinite, the other heavy Higgs
bosons decouple from the theory and cause no problem to the unitarity bound
\cite{susy}.

\section
{Predictions for $\epsilon'/\epsilon$.}

In this section we study the direct CP violation in $K_{L,S}\rightarrow 2\pi$
decays. CP violation in these processes is characterized by the value of
$\epsilon'/ \epsilon$. $\epsilon'/\epsilon$ is defined as
\begin{eqnarray}
{\epsilon'\over \epsilon} = {\omega\xi - ImA_2/ReA_0 \over \xi +
ImM_{12}/\Delta
M_K}\;,
\end{eqnarray}
where $\omega = ReA_2/ReA_0 = 1/20$, $\xi = ImA_0/ReA_0$. Here $A_0$ and $A_2$
are the $\Delta I = 1/2$, $3/2$ decay amplitudes for $K_{L,S}
\rightarrow 2 \pi$.

In our model, the neutral Higgs boson can induce $ImA_i$ at the tree level.
These amplitudes are constrained to be very small due to large Higgs boson
masses.
They are small compared with the CP conserving amplitudes $ReA_i$
generated by W-boson exchange at the tree level. The neutral Higgs boson
contributions to $\epsilon'/\epsilon$ are very small. However there may be
large contributions from the charged Higgs bosons.  The charged Higgs boson
contributions to $ImA_i$ can be generated at the one loop level. The dominant
one is from the charged Higgs boson mediated gluon
penguin. The relevant $\Delta S = 1$ effective Larangian is given by
\begin{eqnarray}
L_{\Delta S=1}= i\bar d \sigma^{\mu\nu}(\tilde f_1{1+\gamma_5\over 2}
+\tilde f_2{1-\gamma_5\over 2})\lambda^a
s G^a_{\mu\nu}\;,
\end{eqnarray}
where $G^a_{\mu\nu}$ is the gluon field strength, and
\begin{eqnarray}
\tilde f_1 &=& {g_s(\mu)\over 32 \pi^2} {m_l\over M^2_{h^+_j}} ({3\over 2}- ln
{m_l^2\over M^2_{h^+_j}})
Im\{(V_{KM}\tilde Y^d_i \gamma_{ij})_{l1}(\tilde Y^{u\dagger}_k
V_{KM}\gamma_{kj})^*_{l2}\}\zeta_f\;,\nonumber\\
\tilde f_2 &=& {g_s(\mu)\over 32 \pi^2} {m_l\over M^2_{h^+_j}} ({3\over 2} - ln
{m_l^2\over M^2_{h^+_j}})Im\{(V_{KM}\tilde Y^d_i \gamma_{ij})^*_{l2}(\tilde
Y^{u\dagger}_k
V_{KM}\gamma_{kj})_{l1}\}\zeta_f\;,
\end{eqnarray}
where $\zeta_f= (\alpha_s(m_h)/\alpha_s(\mu))^{14/23}
 \approx 0.17$ is the QCD correction factor, and l is summed over u, c and t.
We will use
$\alpha_s(\mu) \approx 4\pi/6$ for $\mu = 1 GeV$.
The above effective Lagrangian
will generate a non-zero value
for $ImA_0$\cite{sand}. $L_{\Delta S=1}$ also generates a non-zero
value $\bar \epsilon_{LD}$ for CP violation in
$K^0$ and $\bar K^0$ mixing due to long distance
interactions through $K^0$ and $\pi$, $\eta$, $\eta'$ mixings\cite{dono}. One
obtains\cite{dono,hyc}
\begin{eqnarray}
{\xi\over \bar\epsilon_{LD}}&\approx& -0.196D\;,\nonumber\\
2m_K ImM^K_{12,LD} &\approx& 0.8 \times 10^{-7} (\tilde f_1 + \tilde
f_2)(GeV^3)\;,
\end{eqnarray}
where $D$ is a supression factor
of order $O(m^2_K, m^2_\pi)/\Lambda^2$. ${\xi/ \bar\epsilon_{LD}}$ is of order
-0.014 to -0.1.

We find that in both a) and b) cases, the donimant contributions are
from the top quark in the loop arising from mixing in the charged Higgs boson
couplings. For case a), we have
\begin{eqnarray}
\bar\epsilon_{LD}(h^+_1) &\approx& 18{GeV^2\over m^2_{h^+_1}}
 {m_t\over 150GeV}ln{m_t^2\over m^2_{h^+_1}}\;,\nonumber\\
\bar\epsilon_{LD}(h^+_2) &\approx& 25 {GeV^2\over m^2_{h^+_2}}
{m_t\over 150GeV}ln {m^2_t\over m^2_{h^+_2}}\;,\nonumber\\
\bar\epsilon_{LD}(h^+_3) &\approx& -7 {GeV^2\over m^2_{h^+_3}}
{m_t\over 150GeV}ln {m^2_t\over m^2_{h^+_3}}\;.
\end{eqnarray}
And for case b), we have
\begin{eqnarray}
\bar\epsilon_{LD} \approx -7.35\times 10^{3} Im(\gamma_{22}\gamma_{23}^*){GeV^2
\over
m^2_{\eta^+}}{m_t\over 150GeV}ln{m^2_t\over m^2_{\eta^+}}\;.
\end{eqnarray}
The contributions to $\bar\epsilon$ can be significant in both cases
 depending on the Higgs boson masses and the CP violating parameter
$Im(\gamma_{22}\gamma_{23}^*)$. If the masses of the charged Higgs bosons are
of order
a few hundred GeV, $\bar\epsilon_{LD}$ can be as large as the experimental
value and
$\epsilon'/\epsilon$ can be of order $10^{-3}$. However, there are constraints
 on the charged Higgs boson masses from the experimental upper bound on the
neutron
EDM. We will study these constraints in Sec.VI. When these constraints are
taken into account,
 $\bar\epsilon_{LD}$ is generally constrained to be less
than
$3\times 10^{-5}$ for case a). $\epsilon'/\epsilon$ is then constrained to be
less than
$3\times 10^{-5}$. However, for case b), $\bar\epsilon_{LD}$ can still be as
large as $ 10^{-3}$
and
$\epsilon'/\epsilon$ can be $10^{-3}$.

\section
{ CP violation in the neutral B system.}

There are many processes which can test CP violation in the
neutral B system. Some particularly interesting ones are\cite{bcp}
\begin{eqnarray}
B_d &\rightarrow & J/\psi K_S\;,\nonumber\\
B_d&\rightarrow & \pi^+\pi^-\;,\\
B_s&\rightarrow & \rho K_S\;.\nonumber
\end{eqnarray}
The differences of time variation of decay rates for the above processes and
their
CP tranformed states are given by
\begin{eqnarray}
a_{fCP} &=& {\Gamma (B^0(t) \rightarrow f_{CP}) - \Gamma ( \bar B^0(t)
\rightarrow f_{CP})
\over \Gamma (B^0(t) \rightarrow f_{CP}) + \Gamma ( \bar B^0(t) \rightarrow
f_{CP})}
\nonumber\\
&=& {(1-| \lambda |^2) cos(\Delta M_B t) - 2Im \lambda sin(\Delta M_B t) \over
1 + | \lambda |^2}\;,
\end{eqnarray}
where $f_{CP}$ indicates the final states.  $\lambda$ is defined as
\begin{eqnarray}
\lambda = \left ({q\over p}\right )_B {\bar A \over A} S\;,
\end{eqnarray}
where $(q/p)_B = \sqrt{M^{B*}_{12}/M^B_{12}}$,  $A$ and $\bar A$ are the decay
amplitudes. If the final state contains
$K_S$, $S = (q/p)_K$ which has a phase of order $10^{-3}$. For other cases S is
equal to one.

 Non-zero asymmetry $a_{fCP}$ signals CP violation. If $|\lambda|$ is not equal
to one, it
indicates that CP is violated in the decay amplitudes. In the MSM $|\lambda|$
is
equal to one to a very good approximation for the above three processes.
The asymmetries are proportional to $Im\lambda$. In the MSM, the processes in
Eq.(24) measure three angles related to CP violation in the CKM matrix,
\begin{eqnarray}
Im\lambda (B_d \rightarrow J/\psi K_S) = - sin 2 \beta\;,\nonumber\\
Im\lambda (B_d \rightarrow \pi^+ \pi^-) = sin 2\alpha\;,\\
Im\lambda (B_s \rightarrow \rho K_S) = -sin 2 \gamma\;,\nonumber
\end{eqnarray}
where
\begin{eqnarray}
\alpha &=& arg \left( - {V_{KM,td} V^*_{KM,tb}\over V_{KM,ud}V^*_{KM,ub}}
\right )\;,\nonumber\\
\beta &=& arg \left ( - {V_{KM,cd}V^*_{KM,cb} \over V_{KM,td}V^*_{KM,tb}}
\right )\;,\\
\gamma &=& arg\left (- {V_{KM,ud}V^*_{KM,ub}\over V_{KM,cd}V^*_{KM,cb})}
\right )\;.\nonumber
\end{eqnarray}
The sum of these three angles is equal to $180^0$.

In the $S_3\times Z_3$ model, the situation is very different.
In this model CP violation is mainly due to neutral Higgs boson exchange.
The CP
violating decay amplitudes $A$ and $\bar A$ are small because the decay
amplitudes are dominated by the CP conserving tree level W interactions.
However the phase of $\sqrt{M^{B*}_{12}/M_{12}^B}$ in the $B-\bar B$ mixing due
to neutral Higgs boson exchange can be large.  In case a), there is CP
violation arising from the phase in Yukawa coupling of Higgs bosons, as well as
CKM matrix, but the former is much larger. The three meaurements in Eq.(24) do
not measure the  angles
$\alpha$, $\beta$ and $\gamma$ defined in Eq.(28) anymore. The first two
processes will
mostly measure the phases in $M^{B_d}_{12}$. We have
\begin{eqnarray}
Im\lambda (B_d \rightarrow \pi^+ \pi^-)
 &\approx& Im\lambda (B_d \rightarrow J/\psi K_S) \leq 0.42\;,\;\;from\;
Reh^0_1,\nonumber\\
Im\lambda(B_d \rightarrow \pi^+ \pi^-)
&\approx& Im\lambda(B_d \rightarrow J/\psi K_S) \leq 0.19\;,\;\;
from\;Reh^0_2\;,\\
Im\lambda(B_d\rightarrow \pi^+ \pi^-)
&\approx& Im\lambda(B_d \rightarrow J/\psi K_S)
\approx 0.19\;,\;\;form\;Reh^0_3\;.\nonumber
\end{eqnarray}

For case b),  the CP violation is purely from scalar-pesudoscalar mixing in the
Higgs sector and we find
\begin{eqnarray}
Im\lambda(B_d \rightarrow \pi^+ \pi^-)
\approx Im\lambda(B_d \rightarrow J/\psi K_S) \approx -0.25\;.
\end{eqnarray}

$Im\lambda$ for $B_s \rightarrow \rho K_S$ is different for
a) and b). For case a), the neutral Higgs boson contributions to the asymmetry
are small. However $Im\lambda(B_s \rightarrow \rho K_S)$ due to CP violation in
the KM-matrix can be
about 0.1. For case b), $Im\lambda(B_s \rightarrow \rho K_S)$
from neutral Higgs boson exchange is
only about 0.02.

If interpreted as in Eq.(27), we find for case a), $sin 2\alpha = -sin 2\beta$,
$sin\gamma = 0.05$, and $\alpha + \beta + \gamma \neq 180^0$. For case b),
we have, $sin 2\alpha = -sin 2\beta$, $sin\gamma = 0.01$. We again find,
$\alpha + \beta + \gamma \neq 180^0$. These predictions are different than
those of the MSM.

\section
{ The neutron electric dipole moment.}

The prediction for the neutron EDM in the $S_3\times Z_3$ model is very
different from the MSM. In both a) and b) cases, the neutron EDM can be
generated at the one loop level, and the two loop contributions can be
significant. The one loop contributions in this model are
also different from multi-Higgs models with neutral flavor conservation for
tree level neutral Higgs boson exchange\cite{hyc}. In flavor conserving models
the
fermion EDMs are proportional to the third powers of the external fermion
masses. For u- and d- quarks and electron, the EDMs at the one loop level are
very small ($< 10^{-30} ecm$). In the models considered by us, the couplings
are
quite different from the flavor
conserving models. Furthermore the off diagonal couplings may contribute
significantly. It is possible to have a large neutron EDM at the one loop
level. The d quark EDM due to  neutral Higgs boson exchange at the one loop
level is given by
\begin{eqnarray}
d_d = {Q_d e \over 16\pi^2} Im(a_{i,dl}a_{i,ld}) {m_l\over m_{R_i}^2} ({3\over
2}
-ln \left( {m^2_{R_i}\over m^2_l} \right ))\zeta_d\;,
\end{eqnarray}
where $a_{i,ql} = (\alpha_{ki} + i\alpha'_{ki})\tilde Y^d_{k,ql}$,
 $\zeta_d = (\alpha_s(m_h)/\alpha_s(\mu))^{16/23} \approx 0.12$, and l is
summed over d, s, and b. The quark
EDMs are related to the neutron EDM by quark model,
\begin{eqnarray}
D_n = 4d_d/3 - d_u/3\;.
\end{eqnarray}

For case a) we have
\begin{eqnarray}
D_n(d, Reh^0_1) &=& {4\over 3}d_d(Reh^0_1) \leq 2\times 10^{-28}
ecm\;,\nonumber\\
D_n(d, Reh^0_2) &\leq& 0.7\times 10^{-28} ecm\;,\\
D_n(d, Reh^0_3) &\approx& 2\times 10^{-29} ecm\;.\nonumber
\end{eqnarray}

For case b), we have $D_n(d) \approx 2 \times 10^{-29} ecm$. These values are
at least three orders of magnitude smaller than the experimental upper bound of
$1.2 \times 10^{-25}\; ecm$\cite{nedm}. The u quark EDM is zero at the one loop
level.

In multi-Higgs models, there may be large contributions to
the neutron EDM at the two loop level from the
Weinberg operator\cite{wein} $D_n(W)$ and from the color dipole moment of gluon
due
to Bar-Zee type of diagrams\cite{bz,cdm} $D_n(BZ)$. In our model, we have
\begin{eqnarray}
D_n(W) &\approx& e \zeta_W \Lambda {1\over 64\pi^2} ImZ^i_{tt} {m_t^2\over
m^2_{h^0_i}}ln{m_t^2\over m_{h^0_i}^2}\;,\nonumber\\
D_n(BZ, q) &\approx& {m_q\over 64\pi^3} {c_q\over 9}
\alpha_s(\mu)\zeta_{bz}{m_t^2\over m_{h^0_i}^2}\left ( ln{m_t^2\over
m_{h^0_i}^2}\right )^2
ImZ^i_{tq}\;,
\end{eqnarray}
where $\zeta_W \approx 6\times 10^{-6}$, and $\zeta_{bz} \approx 10^{-2}$
are the QCD
correction factors, $c_u = 2$ and $c_d = 4$, and $\Lambda \approx 1 GeV$ is the
chiral symmetry breaking scale. The parameters $ImZ$ are given by
\begin{eqnarray}
ImZ_{tt}^i &=&{1\over m_t^2} Re(\tilde Y^u_{k, 33} (\alpha_{ki} - i
\alpha'_{ki})) Im(\tilde Y^u_{k',33}(\alpha_{k'i} - i
\alpha'_{k'i}))\;,\nonumber\\
ImZ^i_{tu} &=& {1\over m_u m_t} Im(\tilde Y^u_{k,33} (\alpha_{ki}
-i\alpha'_{ki})
\tilde Y^u_{k',11} (\alpha_{k'i} -i\alpha'_{k'i}))\;\nonumber\\
ImZ^i_{td} &=& {1\over m_d m_t} Im(\tilde Y^u_{k,33} (\alpha_{ki}
-i\alpha'_{ki})
\tilde Y^d_{k',11} (\alpha_{k'i} +i\alpha'_{k'i}))\;.
\end{eqnarray}

For case a), because there is no CP violation in the
up quark sector only down quark loops contribute, $D_n(W)$ from the Weinberg
operator at the
two loop
level is small.
There are non-zero $D_n(BZ)$ from d-quark due to Bar-Zee
mechanism. We
find that the contributions from $Reh^0_{12,}$ is also small ($< 4\times
10^{-28}
ecm$). $Reh^0_3$ contribution is even smaller ($< 10^{-29}$).

For case b), the two loop
contributions to the EDM are significantly larger because in this case there is
CP violation in the top quark interaction. We have
\begin{eqnarray}
D_n(BZ, u) \approx(2 \sim 8)\times 10^{-26} ecm\;,\nonumber\\
D_n(BZ, d) \approx (2 \sim 8)\times 10^{-27} ecm\;,
\end{eqnarray}
for $m_t$ between 100 GeV to 200 GeV. The neutron EDM can be as large as the
experimental upper bound.
The contribution from the Weinberg operator is small, $D_n(W) \leq 10^{-30}
ecm$.

The charged Higgs bosons can also contribute to the neutron EDM. At the one
loop level, the u  and d quark EDM are given by
\begin{eqnarray}
d_u &=& -{1\over 48\pi^2} {m_l \over m^2_{h^+_i}} ln {m^2_l\over m^2_{h^+_i}}Im
[\gamma_{ji}\gamma_{ki}^*(V_{KM}\tilde Y^d_j)_{1l} (V_{KM}^\dagger \tilde
Y^u_k)_{l1}]\;,\nonumber\\
d_d &=& {1\over 24\pi^2}{m_l \over m^2_{h^+_i}} ln {m_l^2\over m^2_{h^+_i}}
Im[\gamma_{ji}\gamma_{ki}^*(V_{KM}\tilde Y^d_j)_{l1} (V_{KM}^\dagger \tilde
Y^u_k)_{1l}]\;.
\end{eqnarray}
For $d_u$, l is summed over d, s, and b; and for $d_d$, l is summed over u, c,
and t.
At the two loop level, there is a large contribution from the Weinberg
operator,
\begin{eqnarray}
D_n(W) \approx e \zeta_W' M {1\over 32\pi^2} ImZ'^i_{tt}{m_t^2\over
m^2_{h^+_i}}
ln{m_t^2\over m^2_{h^+_i}}\;,
\end{eqnarray}
where $\zeta'_W = 3 \times 10^{-4}$ is the QCD correction factor, and
\begin{eqnarray}
ImZ'^i_{tt} = {1\over m_bm_t} Im [ \gamma_{ji}\gamma_{ki}^*(V_{KM} \tilde
Y^d_j)_{33} (V_{KM}\tilde Y^u_k)_{33}]\;.
\end{eqnarray}

We find that in case a) the
dominant contributions are from the two loop Weinberg operator. We
have
\begin{eqnarray}
D_n(W) &\approx &1.6\times 10^{-19} {GeV^2\over m_{h_1^+}^2} ln {m_t^2\over
m_{h_1^+}^2}{m_t^2\over (150 GeV)^2}\;ecm\;,\nonumber\\
D_n(W) &\approx& 1.4\times 10^{-19} {GeV^2\over m_{h_2^+}^2} ln {m_t^2\over
m_{h_2^+}^2}{m_t^2\over (150 GeV)^2}\;ecm\;,\nonumber\\
D_n(W) &\approx& 1.2\times 10^{-25} {GeV^2\over m_{h_3^+}^2} ln {m_t^2\over
m_{h_3^+}^2}{m_t^2\over (150 GeV)^2}\;ecm\;.
\end{eqnarray}
Requiring the contributions to be less than the experimental value, we find
the masses of $h^+_{1,2}$ have to be larger than $2.5 TeV$. There is no
constraint on $h^+_3$ mass. Combining this information with those from Eqs.(22)
and (23),
we find the charged Higgs boson contributions to $\bar \epsilon_{LD}$ is less
than
$3\times 10^{-5}$, and $\epsilon'/ \epsilon$ is
less than $3 \times 10^{-5}$.

For case b), we find the dominant contribution is from the one loop d quark
EDM.
We have
\begin{eqnarray}
D_n(d) \approx 5.4 \times 10^{-19} Im(\gamma_{22}\gamma_{23}^*){GeV^2\over
m^2_{\eta^+}}
ln{m_t^2\over m_{\eta^+}^2}{m_t\over 150GeV}\; ecm\;.
\end{eqnarray}
Requiring $D_n(d)$ to be less than the experimental value,
$\bar \epsilon_{LD}$ is constrained to be less than $10^{-3}$, and $\epsilon'/
\epsilon$ can still be of order $10^{-3}$. Assuming maximum mixing, the mass
of $\eta^+$ is constrained to be larger than $5\; TeV$.

We also checked cases for different values of $\sigma$ and different mixings
between $Reh_i$ and $Imh_j$. We find that for case a) if $95^0 >|\sigma| >
85^0$, it is
not possible to produce the experimental value for $\bar\epsilon$ because the
constraints on the neutral Higgs boson masses from the mass difference of the
neutral B
mesons is too strong. The model can explain the observed CP violation in the
neutral K system even for $|\sigma| = 10^0$, and $|180^0-\sigma| = 10^0$.
The neutral Higgs boson masses are typically constrained to
be
larger than a few TeV. For angles close to zero and $180^0$, CP violation for B
system become
small. $Im\lambda (B_d \rightarrow J/\psi K_S)\approx Im \lambda (B_d
\rightarrow \pi^+\pi^-)$ are between $0.05 \sim 0.45$. $Im(B_s \rightarrow \rho
K_S)$ is less than 0.1. The neutron EDM due to the neutral Higgs bosons is
typically less than $10^{-27}\; ecm$. The contributions
to the neutron EDM from the charged Higgs bosons can be close to the
experimental upper bound. $\epsilon'/\epsilon$ are typically less than
$10^{-4}$.
 For case b),  $Im\lambda(B_d \rightarrow \pi^+\pi^-)$ and $Im\lambda(B_d
\rightarrow J/\psi
K_S)$ vary between 0.02 to 0.3. $Im\lambda (B_s\rightarrow \rho K_S)$ is less
than 0.1. The neutron EDM due to neutral Higgs bosons is larger than $10^{-27}
edm$
and can be as large as the experimental upper bound. The charged Higgs
bosons contribution to the neutron EDM can be close to the experimental upper
bound. $\epsilon'/\epsilon$ can be of order $10^{-3}$.

The predictions also depend on the choices of the VEVs. The general features
are, however,  the same. For example for
$v_1 =210\; GeV$, $v_2 = 3; GeV$, $v_3 = 0.5\;GeV$ and $v_4 = 130 \;GeV$, we
find:  1)The neutral Higgs bosons are constrained to be in the multi TeV
region;
2) $\epsilon'/\epsilon$ is small in case a) and can be $10^{-3}$ in case
b); 3) the predictions for CP violation B system maintain the same features;
and 4) the predictions for the neutral EDM from the neutral Higgs bosons are in
the same regions as discussed earlier.

\section{ The electron electric dipole moment}

The $S_3\times Z_3$ model may also have interesting CP violating signatures in
the lepton sector.  We assume the same $S_3\times Z_3$
assignments for the left handed and the charged right handed leptons as their
quark partnenrs\cite{ma}. The mass matrix and Yukawa couplings for the charged
leptons
are similar to the down quarks. One simply changes the  parameters (a, b, c, d
, and $\xi$) for quarks to ($a_l\;, b_l\;, c_l\;, d_l\;,$ and $\xi_l =
|\xi|e^{i\sigma'}$) for leptons. We use\cite{desh}: $a_l=0.106 GeV\;, b_l=0\;,
c_l=1.781
GeV\;, d_l = 8.6\times 10^{-3} GeV$. For this set of parameters, we have $m_e =
0.511 MeV$, $m_\mu = 106 MeV$ and $m_\tau = 1784 MeV$ which are in good
agreement with experimental data. We again consider two cases: a) CP violation
purely due to complex Yukawa coupling with the phase $\sigma' =80^0$; and case
b) CP violation due to $Reh^0_2$ and
$Imh^0_3$ mixing. If right handed neutrinos exist, the charged Higgs bosons
will
also contribute to electron EDM. However due to very small neutrino masses, the
contribution to the electron EDM is neglegiblly small. We will not consider
them
here.

For case a) we find that the one loop contributions are small ($< 10^{29}
ecm$). However the two loop contribution due to Bar-Zee mechanism\cite{bz,edm}
can be as large as
\begin{eqnarray}
d_e (Reh^0_1) \leq  10^{-27} ecm\;,\nonumber\\
d_e (Reh^0_2) \leq 1.5 \times 10^{-27}\;
\end{eqnarray}
for $m_t < 200 GeV$.
$Reh^0_3$ contribution is much smaller.
For case b), we find that the one loop and two loop contributions are small
 ($< 10^{-33}ecm$).

For case a) the predictions depend on the phase $\sigma'$. However as
long as the phase is
not too close to $0^0$ and $180^0$ ($|sin\sigma'| \geq 0.17$), the electron EDM
can be a few times of
$10^{-27}\; ecm$. It is below the experimental upper bound\cite{eedm}. For case
b), different mixings may have different values
for electron EDM and can be larger than the special case considered here, but
is
less than $10^{-27}\; ecm$. For
example, the case for $Re^0_1$ and $Imh^0_2$, $d_e$ is about $2\times
10^{-28}\; ecm$.
Varying VEVs can change the predictions for the electron EDM. Using the set of
VEVs discussed at the end of the last section, we find the value for electron
EDM is
smaller by one order of magnitude for case a), and for case b) we again obtain
small electron EDM.

\section
{ Conclusions}

We have studied in detail some effects due to two different CP violating
mechanisms in the $S_3\times Z_3$ model. Both mechanisms discussed in this
paper can explain the observed CP violation in the neutral K system.
CP violation in the neutral K system
and the mass difference in the neutral B system constrain the neutral Higgs
boson masses to be in the multi TeV region. The predictions for other CP
violations observables are very different from the MSM.

i) \underline{$\epsilon'/\epsilon$} : In the MSM, depending on the top quark
mass, $\epsilon'/\epsilon$ can be as
large as $10^{-3}$. In the $S_3\times Z_3$ model $\epsilon'/ \epsilon$ due to
neutral Higgs bosons is small. The charged Higgs boson cantribution for
$\epsilon'/\epsilon$ in case
a) is also small. But for case b), $\epsilon'/\epsilon$ can be as large as
$10^{-3}$. The measurement of $\epsilon'/\epsilon$ may distiguish the MSM and
case b) from case a).

ii) \underline{CP violation in B system}: In section V, we discussed three
asymmetries in the neutral B system. In the MSM model these three asymmetries
measure three angles related to the CKM matrix. The sum of these angles is
equal to $180^0$. However in the $S_3\times Z_3$ model, CP violation in the
neutral B system comes from different sources. The predictions are very
different from the MSM. If we interpret the measurements of the three
parameters in Eq. (27) in terms of the three angles, we find their sums are not
equal to $180^0$ in both a) and b) cases and more over $sin2\alpha =
-sin2\beta$.  These experiments should distinguish the MSM from the $S_3\times
Z_3$ model and the B factory will provide us with very useful
information.

iii) \underline{The Electric Dipole Moment}: The predictions for the neutron
and electron EDMs are
several orders of magnitude larger than those from the MSM. There are also
dramatic differences between the two cases considered. The neutron EDM can be
as large as the experimental upper bound for case b). It is smaller in case a).
The electron EDM is below the experimental bound, but future experiments will
reach the sensitivity necessary to test case a). The electron EDM is smaller in
case b). It is also interesting to note
that some of the charged Higgs boson masses are also constrained to be in the
TeV region. This is different from flavor conserving multi-Higgs models, where
limits are much weaker.

\acknowledgments
This research was supported by the Department of Energy Grant No.
DE-FG06-85ER40224.

\end{document}